\documentclass[12pt,a4paper, amsmath,amssymb]{article}
\usepackage[margin=0.8in]{geometry}
\usepackage{amssymb}
\usepackage{amsmath}
\usepackage{amsfonts}
\usepackage{bm}
\usepackage[utf8]{inputenc}
\usepackage{graphicx}
\usepackage{float}
\usepackage[colorlinks,linkcolor = blue,
urlcolor  = blue,
citecolor = blue,
anchorcolor = blue]{hyperref}
\usepackage{xcolor}
\usepackage{ctable} 
\usepackage{adjustbox}
\def\thefootnote{\fnsymbol{footnote}}
\usepackage{cite}
\begin{document}

\begin{center}
{\Large \textbf{Thermodynamic Features of AdS Black Holes within the Rastall Gravity and Perfect Fluid Matter Framework}}
\thispagestyle{empty}

\vspace{1cm}

{\sc

H. Laassiri$^1$\footnote{\url{hayat.laassiri@gmail.com}},
A. Daassou$^1$\footnote{\url{ahmed.daassou@uca.ma}},
R. Benbrik$^1$\footnote{\url{r.benbrik@uca.ac.ma}}\\
}

\vspace{1cm}

{\sl
  $^1$Fundamental and Applied Physics Laboratory, Physics Department, Polydisciplinary Faculty,
  Cadi Ayyad University, Sidi Bouzid, B.P. 4162, Safi, Morocco.\\

}

\end{center}

\vspace*{0.1cm}

\begin{abstract}
In this study, we  analyzed the impact of a perfect fluid on the phase transition of Anti-de Sitter (AdS) black holes  within the Rastall gravitational background.  Compared to similar studies in the  literature,  the findings of this work  are highlighted by the determination of analytical expressions for critical points for charged and Kerr-Newman AdS black holes  using approximated formulas for the horizon radius.  An accurate analysis of these new analytical expressions  allowed us to discover a new
viable condition that relates the Rastall parameter $\kappa\lambda$ to the equation of state parameter $\omega$,   expressed as: $\kappa\lambda = \frac{\omega}{1 + \omega}$.  Thanks to this new condition, we were able to reproduce all the analytical expressions of critical points calculated within the framework of Einstein’s general relativity for two cases: a charged AdS black hole and a rotating AdS black hole. These findings suggest that Rastall gravity,   considering this new condition, could serve as an alternative theory of gravitation to general relativity. The  approximate  expression  of the horizon radius also  enabled  the exploration of  the distinctiveness of fractional-order phase transitions in these AdS black holes. Furthermore, we calculated the critical exponents, offering insights into the behavior of crucial thermodynamic quantities near the inflection point.  Examining  how a perfect fluid influences phase transition reveals various critical behaviors, demonstrating that the variation in   the phase transition  depends on the intensity of the perfect fluid. Notably, this variation is portrayed by a linearly increasing trajectory with the escalation of this intensity.

\end{abstract}

 Keywords: Perfect fluid matter; Critical phenomena; Ads black holes; Rastall gravity.

\def\thefootnote{\arabic{footnote}}
\setcounter{page}{0}
\setcounter{footnote}{0}

\newpage
\section{Introduction}
\label{intro}\
P. Rastall proposed Rastall gravity in 1972, postulating that the covariant derivative of the energy-momentum tensor should be zero within a flat spacetime framework~\cite{1}. A significant characteristic of this theory is its capability to account for the accelerated expansion of the cosmos~\cite{2}.   Additionally, this theory incorporates the process of gravitational collapse in a homogeneous perfect fluid~\cite{3}, and it offers  the potential for the existence of traversable wormhole solutions with positive feasibility~\cite{4}.   Recently, the investigation of black holes within the framework of Rastall gravity has garnered significant interest, leading to a diverse range of research in the literature. This research includes various phenomenological outcomes, particularly those associated with cosmological aspects~\cite{5,6,7,8,9,10,11}, and the implications it  holds for astrophysical scenarios~\cite{12,13,14,15,16}.\

The exploration of black hole thermodynamics remains a pertinent and essential subject within the realm of theoretical physics~\cite{17,18}.  Recently, there has been a substantial upsurge in the study of phase transitions in black holes situated in asymptotically Anti-de Sitter (AdS) space. This  increased interest is  mainly due to the association of these transitions with holographic superconductivity within the AdS/CFT correspondence framework \cite{19,20,21,22,23}. Numerous investigations have highlighted that charged AdS black holes exhibit phase transitions between small and large configurations in the canonical ensemble with a constant charge. Interestingly, these transitions draw an analogy with the liquid-to-gas phase transitions observed in van der Waals (vdW) fluids~\cite{24,25,26}. Comparable patterns were subsequently identified in the case of Kerr-AdS and Kerr-Newman-AdS black holes~\cite{27,28,29,30}. Precisely establishing an analogy between an AdS black hole system and a van der Waals (vdW) fluid was achieved by treating the cosmological constant as a thermodynamic pressure and its corresponding conjugate as a volume ~\cite{31,32,33,34}.\

A multitude of studies have highlighted the existence of phase transitions involving the transformation between small and large black holes within AdS black hole systems~\cite{35,36,37,38,39,40,41,42,43,44,45,46,47,48,49,50,51,52,53}. Because these systems demonstrate analogous behavior to a van der Waals (vdW) fluid, they display similar features, such as the swallowtail pattern in Gibbs free energy, oscillatory behavior in isothermal lines, and matching scaling laws and critical exponents in the vicinity of the critical point. Moreover, in Refs.~\cite{54,55,56,57,58,59,60},  multi-critical occurrences, such as the triple point and the phase transition with reentrant behavior, were also noted. In Ref.~\cite{61}, an innovative approach to drawing an analogy between AdS black holes and vdW fluids was proposed. This involved the introduction of the concept of black hole  number density to explain the phase transition between small and large black holes. The existence of interaction between two
black hole molecules was also revealed. These findings offer a novel viewpoint on the phase transition of AdS black holes. In line with the proposal of Ref.~\cite{53}, the Reissner-Nordström (RN)-AdS and Kerr-AdS black holes represent thermodynamic systems with a singular characteristic parameter.
  A Kerr-Newman-AdS black hole can be described as a thermodynamic system characterized by two distinct parameters. Consequently, the critical points are indeed dependent upon the values of $ Q $ and $ j $. Nevertheless, the reduced temperature and Gibbs free energy solely rely on the dimensionless ratio of angular momentum-charge (AMC) $ j/Q ^{2} $ rather than $ j $ and $ Q $~\cite{62}.\

In Ref.~\cite{63}, the authors conducted an investigation into the phase transition of charged AdS black holes within Gauss-Bonnet gravity. They concluded that when quintessence is present, the charged AdS black hole exhibits a phase transition from small to large configurations, resembling the liquid-to-gas transition observed in van der Waals (vdW) fluids. Our recent work, Ref.~\cite{64}, reveals that when a cloud of strings and quintessence surrounds a charged rotating AdS black hole, these two additional sources of energy have no effect on the existence of the small-large black-hole phase transition. The influence of dark energy on the P-V critical behavior of RN-AdS black holes was investigated in Ref.~\cite{65}. The study revealed that the presence of quintessence matter does not alter the occurrence of the small-large black hole phase transition. This was determined by applying the Maxwell equal-area law. The authors of Ref.~\cite{66} analyzed the phase transition of charged AdS black holes while considering the presence of quintessence matter. The results indicate that these black holes undergo a phase transition analogous to that observed in van der Waals (vdW) fluids. In Refs.~\cite{67,68,69}, the authors examined the phase transition of a quintessential Kerr-Newman-AdS black hole using the Maxwell equal-area law. The study's conclusion suggests that below the critical temperature, the black holes exhibit a phase transition resembling that of the van der Waals (vdW) system.

This paper focuses on exploring how the inclusion of perfect fluid matter affects the critical behavior and phase transitions  for AdS black holes  within the Rastall gravity framework.  We aim to examine the impact  of perfect fluid matter  on  the critical phenomena and phase transitions  of AdS black holes  within the wider framework of Rastall gravity theory.   The paper is  organized as follows. Section \ref{Sec:2}  
provides an overview of the thermodynamic characteristics of AdS black holes in the presence of perfect fluid matter   within the   framework of Rastall gravity. 
In Section \ref{Sec:3}, we   derive analytical expressions for the critical points associated with charged and Kerr-Newman Anti-de Sitter (AdS) black holes. These equations  account for factors such as the intensity of perfect fluid matter,  two system-specific parameters ($Q$ and $j$), Rastall theory, and equation of state parameters. This analysis is  performed within the context of approximate expressions for horizon radius values, applying critical criteria in the process.  In Section \ref{Sec:4}, we compute critical exponents to elucidate the behavior of crucial thermodynamic quantities near the inflection point.
Section \ref{Sec:5}   explores the impact of perfect fluid matter on the phase transition of charged rotating Anti-de Sitter (AdS) black holes, emphasizing  various critical behaviors and  highlighting  the significance of perfect fluid intensity in this process. Moving forward to Section \ref{Sec:6},  we calculate the   fractional   order  phase  transitions of these black holes.  Finally, Section \ref{Sec:7}  offers a comprehensive summary and discussion of the  achieved results.
\section{Thermodynamic properties of  AdS black holes surrounded by perfect fluid matter within the framework of Rastall gravity}
\label{Sec:2}

The metric describing a charged rotating  Anti-de Sitter (AdS) black hole in the presence of perfect fluid matter within the framework of Rastall gravity, incorporating the influence of the cosmological constant, is provided in \cite{70}:

\begin{equation}\label{Eq.1}
\begin{aligned}
\text{ds}^2 &= \frac{\Sigma ^2}{\Delta _r}\text{dr}^2+\frac{\Sigma ^2}{\Delta _{\theta }}\text{d$\theta $}^2+\frac{\Delta _{\theta
	}\sin^2\theta }{\Sigma ^2}\left(a \frac{\text{dt}}{\Xi }-\left(r^2+a^2\right)\frac{\text{d$\phi $}}{\Xi }\right)^2\
-\frac{\Delta _r}{\Sigma
	^2}\left(\frac{\text{dt}}{\Xi }-a \sin^2\frac{\text{d$\phi $}}{\Xi }\right)^2,
\end{aligned}
\end{equation}

where
\begin{equation}\label{Eq.2}
\begin{aligned}
\Delta _r &= r^2-2\text{mr}+a^2+q^2-\alpha r^{n}-\frac{\Lambda }{3}r^2\left(r^2+a^2\right), \quad \text{with} \quad n = \dfrac{1-3\omega}{1-3\kappa \lambda (1+\omega )},
\end{aligned}
\end{equation}

\begin{equation}\label{Eq.3}
\begin{aligned}
\Delta _{\theta } &= 1+\frac{\Lambda }{3}a^2\cos^2\theta, \quad \Xi = 1+ \frac{\Lambda }{3}a^2.
\end{aligned}
\end{equation}

In this context, the symbol $ \omega $ characterizes the equation of state, while $ \alpha $  denotes the intensity level  of the perfect fluid matter surrounding the black hole. The term $ \kappa\lambda $  represents the Rastall parameter,  with $ m $ and $ q $ represent the mass and charge  parameters of the black hole, respectively. Regarding the equation of state \cite{70}, represented by $ \omega $, when $ \omega $ falls within the range of $ -1 $ to $ -1/3 $, it implies that the perfect fluid substance exhibits characteristics of dark energy. Conversely, if $ \omega $ is precisely $ -1/3 $, it signifies that the perfect fluid substance embodies the properties of perfect fluid dark matter. When examining Eq. (\ref{Eq.4}) of the energy condition, for dark energy with $\omega = -\frac{2}{3}$, it results in the constraint $-2 \leq \kappa\lambda \leq \frac{1}{4}$. Similarly, for dark matter with $\omega = -\frac{1}{3}$, the constraint becomes $-\frac{1}{2} \leq \kappa\lambda \leq \frac{1}{4}$. By utilizing Eq. (\ref{Eq.2})   and assuming the cosmological constant $ \Lambda $ is negligible (set to 0 due to its small value), especially for perfect fluid dark matter with $ \omega = -1/3 $, calculations demonstrate that $ \alpha $ must adhere to the condition: $ 0 < \alpha \leq \dfrac{2^{\frac{\kappa \lambda }{\kappa \lambda -1}}(1-\kappa \lambda )^2}{3(2+\kappa \lambda )} $.  Additional  scenarios can  be explored using a similar methodology for various $ \omega $ values. These findings indicate that $ \kappa \lambda $, governing the redistribution of perfect fluid  matter, and the equation of state $ \omega $ impose constraints on $ \alpha $ within the framework of Rastall gravity.

The weak energy condition (WEC) mandates non-negativity of the total energy density for any timelike observer, $T_{\mu\nu}u^{\mu}u^{\nu} \geq 0$, where $u^{\nu}$ represents the timelike vector, and $T_{\mu\nu}$ stands for the diagonal energy-momentum tensor. The strong energy condition (SEC) requires satisfaction of the Raychaudhuri equation, $(T_{\mu\nu}-\frac{1}{2}Tg_{\mu\nu})u^{\mu}u^{\nu} \geq 0$. It's   worth noting that when the condition

\begin{equation}\label{Eq.4}
(-3 \omega +3 \kappa \lambda (1+\omega ))\alpha (1-4 \kappa \lambda )\geq 0,
\end{equation}
is fulfilled, both the WEC and SEC   hold true for the KN-AdS black hole.\\

The expression for the Hawking temperature of these specific black hole types is as follows
\begin{equation}\label{Eq.5}
\begin{aligned}
{T= \dfrac{\Delta'(r)|_{r=r_{h}}}{4\pi (r_h^2 + a^2)} =-\dfrac{a^2 \left(3+\Lambda  r_h^2\right)}{12 \Xi  r_h}+\dfrac{-Q^2 \Xi ^2+r_h^2-\Lambda  r_h^4+\dfrac{3 \alpha  (\kappa \lambda -\omega
			+\kappa \lambda  \omega ) r_h^n}{-1+3 \kappa \lambda  (1+\omega )}}{4 \Xi  r_h}}.
		\end{aligned}
\end{equation}
 In the usual interpretation, we  consider the cosmological constant as a dynamic pressure characteristic of the black hole~\cite{32,34},   with  its corresponding conjugate  representing  physical volume.

\begin{equation}\label{Eq.6}
P=- \frac{\Lambda}{8 \pi},\hspace{1cm}{V=\dfrac{\partial M}{\partial P} ={\dfrac{\pi v^3}{6}}},
\end{equation}
Here, $v$ represents the specific volume.

The black hole horizon $r_h$ is determined from the condition $ \Delta _{r} = 0 $. The black hole's entropy $S$ is formulated as follows~\cite{29}:

\begin{equation}\label{Eq.7}
{S=\frac{\pi \left(a^2+r_h^2\right)}{\Xi }}.
\end{equation}	
The physical mass $M$, charge $Q$, and angular momentum $j$ of the black hole can be described using the parameters $m$, $q$, and $a$.

\begin{equation}\label{Eq.8}
{M = \dfrac{m}{\Xi ^2}}.\hspace{2cm}{Q = \dfrac{q}{\Xi }}.\hspace{2cm}{j= \dfrac{a m}{\Xi ^2}}.
\end{equation}

In classical thermodynamics, the mass $ M $ of the black hole can be understood as the enthalpy $ H $, rather than representing the total energy of the spacetime~\cite{31,54}. Consequently, based on the equations provided earlier, the Gibbs free energy is formulated as follows:
\begin{equation}\label{Eq.9}
G = M-TS,
\end{equation}
\begin{equation}\label{Eq.10}
\begin{aligned}
G = \dfrac{(2+\Xi ) \left(a^2+Q^2 \Xi ^2\right)}{4 \Xi ^2 r_h}+\dfrac{\left(-3+a^2 \Lambda \right) (-2+\Xi ) r_h}{12 \Xi ^2}+\dfrac{\Lambda
		(-2+3 \Xi ) r_h^3}{12 \Xi ^2}\\
-{\dfrac{\alpha  (-2-3 \Xi  \omega +3 \kappa \lambda  (2+\Xi ) (1+\omega )) }{4 \Xi ^2 (-1+3 \kappa \lambda  (1+\omega ))}}{r_h^{-\dfrac{3 (\kappa \lambda -\omega +\kappa \lambda  \omega )}{-1+3 \kappa \lambda  (1+\omega )}}}.
	\end{aligned}
\end{equation}
The heat capacity can be elaborated on as follows:

\begin{equation}\label{Eq.11}
\begin{aligned}
C_{P,j,Q,\alpha ,\kappa \lambda,\omega } = T {\left(\frac{\partial S}{\partial T}\right)_{P,j,Q,\alpha ,\kappa \lambda ,\omega}}.
\end{aligned}
\end{equation}

It's essential to highlight that the heat capacity signifies a condition of local thermodynamic stability. Positive and negative heat capacities indicate stable and unstable systems, respectively. Additionally, it's established that the occurrence of a phase transition coincides with the point of divergence in the heat capacity.

\section{   Critical analysis of AdS black holes in Rastall gravity and perfect fluid matter}
\label{Sec:3}
In this section,   our objective is to    develop analytical expressions for black holes  in the presence of perfect fluid matter and  within the framework of Rastall gravity.  Obtaining an analytical expression for the critical point  involves solving a system of equations derived from the critical condition, which based on the expression of the Hawking temperature Eq. (\ref{Eq.5}). The two equations of this system are  undoubtedly of arbitrary order, owing to the presence of the term  $ r^n $ in the expression of  $T$,   which makes exact resolution impossible without resorting to approximations. In our approach, we  used  the Taylor series to express the horizon, as  shown in Eq. (\ref{Eq.12}), to make the mathematical problem tractable. This approximation is  necessary to effectively address the problem of determining the critical point.  We chose to truncate the series at the first order, as it   provided results with minimal relative deviation  compared to charged rotating AdS black holes   without considering additional parameters in these calculations.  Moreover, limiting the order to 1 helps avoid unnecessary complications.

\begin{align}\label{Eq.12}
r^{n} = c^{-1+n} (c-c n+n r)+O[r-c]^2, 
	\end{align}	
with $c$ representing the convergence value of the horizon radius, substituting the expressions from Eqs. (\ref{Eq.6})  and (\ref{Eq.12}) into Eq. (\ref{Eq.5}), we obtain the resulting expression for the Hawking temperature:
\begin{align}\label{Eq.13}
T = \dfrac{-64a^4P^2 \pi^2 Q^2 + 3a^2 \left(-3 + 8P\pi \left(2Q^2 + r_{h}^2\right)\right) + 9 \left(-Q^2 + r_{h}^2 + 8P\pi r_{h}^4 - c^n (-1+n) \alpha \right)}{36\pi r_{h} \left(a^2 + r_{h}^2\right)}.
\end{align}	
The scenario of charged AdS black holes with perfect fluid  in Rastall gravity is achieved by setting $a=0$. Solving the system of two equations derived from the critical condition 
\[
\frac{\partial T}{\partial r_{h}}\bigg|_{\kappa \lambda, \alpha, Q, \omega, P} = \frac{\partial ^2T}{\partial ^2r_{h}}\bigg|_{\kappa \lambda, \alpha, Q, \omega, P} = 0,
\]
provides us with the analytical expression of the critical point.\
\begin{equation}\label{Eq.14}
\begin{aligned}
P_c &= \dfrac{1}{96 \pi \left(Q^2 - c^n \alpha + c^n n \alpha \right)}, \\
v_c &= 2\sqrt{6} \sqrt{Q^2 - c^n \alpha + c^n n \alpha}, \\
T_c &= \dfrac{\sqrt{6}}{18\pi \sqrt{Q^2 + c^n (-1+n) \alpha}}, \\
r_c &= \sqrt{6} \sqrt{Q^2 - c^n \alpha + c^n n \alpha},\\
G_c &= \dfrac{4 c Q^2 + 4 c^{1+n} (-1+n) \alpha - \sqrt{6} c^n n \alpha \sqrt{Q^2 + c^n (-1+n) \alpha}}{2 \sqrt{6} c \sqrt{Q^2 + c^n (-1+n) \alpha}}.
\end{aligned}
\end{equation} 			
When dealing with charged AdS black holes  and assuming $\kappa\lambda$ and $\alpha$ are both held constant at zero, the crucial thermodynamic property values determined through this investigation are as  follows:
\[ T_c = \frac{\sqrt{6}}{18\pi Q}, \quad S_c = 6\pi Q^2, \quad P_c = \frac{1}{96\pi Q^2}, \quad v_c = 2\sqrt{6} Q, \quad G_c = \frac{\sqrt{6}}{3} Q.\]
It's worth emphasizing that these equations correspond to the expressions observed in charged AdS black holes~\cite{34}. This finding validates the soundness of the initial approximation applied.
 			
 		 Following the provided procedure, we derive the analytical expression for the critical point in Kerr-Newman AdS black holes with perfect fluid within the framework of Rastall gravity. Eq. (\ref{Eq.7}) yields the subsequent expression for the horizon radius \
 		\begin{align}\label{Eq.15}
{r_{h} = \sqrt{\frac{3S-\pi  a^2 (3+8 P S)}{3\pi }}},
 		\end{align}
 		By employing Eq. (\ref{Eq.8}) and Eq. (\ref{Eq.15}), and expanding all intricacies up to  \( O(j^2) \), we achieve
 		\begin{align}\label{Eq.16}
{a= \dfrac{6 c j \sqrt{\pi } \sqrt{S}}{3 c \pi  Q^2+3 c S+8 c P S^2-3 c^{1+n} \pi  \alpha +3 c^{1+n} n \pi  \alpha -3 c^n n \sqrt{\pi
 				} \sqrt{S} \alpha }+ O(j^{2}).}
 		\end{align}	
 		Substituting Eq. (\ref{Eq.15}) and Eq. (\ref{Eq.16}) into Eq. (\ref{Eq.13}) and applying the critical condition ${\dfrac{\partial T}{\partial S}|_{\kappa \lambda, j, \alpha, Q, \omega ,P}} = \dfrac{\partial ^2T}{\partial ^2S}|_{\kappa \lambda, j, \alpha, Q, \omega, P} = 0$,  and  after simplifying lengthy and intricate calculations, the analytical expression for the critical point is as follows:\  
\begin{equation}\label{Eq.17}
\begin{aligned}
P_c &= \dfrac{5 j^2 \left(2 \gamma_1 - 3 \gamma_2\right) + \left(Q^2 + c^n (-1 + n) \alpha \right)^2 \left(\gamma_1 - \gamma_2\right)}{3600 j^4 \pi }, \\
T_c &= \left(-2 j^2 150 j^4 \pi + \left(\gamma_1 + \gamma_2\right) \left(-\pi \left(Q^2 + c^n (-1 + n) \alpha \right)150 j^4 + 3 \pi \left(\gamma_1 + \gamma_2\right) \left(150 j^4 + \left(5 j^2 \left(2 \gamma_1\right.\right.\right.\right.\right. \\
&\left.\left.\left.\left.\left.- 3 \gamma_2\right) + \left(Q^2 + c^n (-1 + n) \alpha \right)^2 \left(\gamma_1 - \gamma_2\right)\right) \left(\gamma_1 + \gamma_2\right)\right)\right)\right)/\left(150 j^4 12 \sqrt{3} \pi^2 \left(\gamma_1 + \gamma_2\right)^{5/2}\right), \\
S_c &= 3 \pi \left(\gamma_1 + \gamma_2\right), \\
G_c &= -T S_c + \left(\left(3 c \pi \left(Q^2 + c^n (-1 + n) \alpha \right) - 3 c^n n \sqrt{\pi} \alpha \sqrt{S_c} + 3 c S_c + 8 c P_c S_c^2\right)^3 + 6 c^2 j^2 \pi^2 \right. \\
&\left. \left(-96 c^n n \sqrt{\pi} \alpha P_c S_c^{3/2} + 3 c^{1+n} (-1 + n) \pi \alpha \left(3 + 40 P_c S_c\right) + c \left(3 + 8 P_c S_c\right) \left(3 \pi Q^2 + S_c \left(3 + 8 P_c S_c\right)\right)\right)\right) \\
&\left/ \left(6 c \sqrt{\pi} \sqrt{S_c} \left(3 c \pi \left(Q^2 + c^n (-1 + n) \alpha \right) - 3 c^n n \sqrt{\pi} \alpha \sqrt{S_c} + 3 c S_c + 8 c P_c S_c^2\right)^2\right)\right., \\
\end{aligned}
\end{equation}
with $\gamma _1=\sqrt{10 j^2+\left(Q^2+c^n (-1+n) \alpha \right)^2}$ and $\gamma _2=Q^2-c^n \alpha +c^n n \alpha$.

When setting $\kappa\lambda = \alpha = Q = 0$, we derive the following:
\begin{align*}
P_c &= \dfrac{1}{12j\sqrt{90}{ \pi}}, \hspace{2cm} T_c = \dfrac{90^{3/4}}{225\sqrt{j}{\pi}}.
\end{align*}
We obtain the same approximate expressions as those cited in Ref.~\cite{71},  enhancing the reliability and consistency of  our approach   in deriving critical expressions for rotating AdS black holes influenced by perfect fluid within the framework of Rastall gravity.

We define the relative deviation $\Delta X_c = \frac{X_c - X_{ac}}{X_c}$ for a thermodynamic quantity $X_c$, where $X_{ac}$ represents the critical point obtained from our work for rotating AdS black holes, with  additional parameters $(\kappa, \lambda, \alpha, \omega)$ set to zero,  considering the analytical result of rotating AdS black holes provided in Ref.~\cite{53}. Under these conditions, the relative deviation in thermodynamic quantities can be expressed as follows:
\begin{align*}
\Delta T_c &= 0.98\%,\hspace{1cm} \Delta G_c = -0.75\%.
\end{align*}
We can observe that the discrepancies between the analytical and approximate values of the critical point are very small. This corresponds to the deviations noticed between the analytical and approximate values for the $d = 4$-dimensional Kerr-AdS black hole, as detailed in Ref.~\cite{53}.   These deviations can be attributed to the difference in applicability: the analytical critical points are valid for all $j$ values, whereas the approximated critical points derived from this study are specifically   calculated under the assumption of small $j$ values.

 In our analysis,  we uncover  a new  relationship between the Rastall parameter $\kappa\lambda$  and the equation of state parameter $\omega$: $\kappa\lambda = \dfrac{\omega}{1 + \omega}$.    This discovery allows us to replicate all the analytical expressions of critical points   derived within the framework of Einstein's general relativity for    the two cases: a charged AdS black hole and a rotating AdS black hole. These derivations are conducted within the context of Rastall gravity  coupled with perfect fluid.  Our findings suggest that Rastall gravity,  due to this newly revealed condition,  offers a credible alternative to general relativity in gravitational theories.
\section{ Response around the inflection point }
\label{Sec:4}
We will compute the critical exponents $\gamma$, $\delta$, and $\beta$ for an AdS black hole surrounded by perfect fluid matter in the context of Rastall gravity. These critical exponents play a significant role in describing different aspects of the system: $\gamma$ characterizes the behavior of isothermal compressibility, $\delta$ pertains to the behavior of the critical isotherm in relation to the critical temperature $T_c$, and $\beta$ delineates the behavior of the order parameter. To facilitate these calculations, we have adopted the following notations,
\begin{align*}
p &= \frac{P}{P_c}, & \epsilon &= \frac{v}{v_c} - 1, & t &= \frac{T}{T_c} - 1.
\end{align*}
The variables $p$, $\epsilon$, and $t$ represent the reduced pressure, volume, and temperature, respectively. These scaled quantities enable the derivation of dimensionless parameters and equations that describe the system's behavior near its critical point.

The critical exponents can be expressed in terms of a power-law relationship around the critical temperature, denoted as follows:
\begin{align*}
 C_v &\propto \vert t \vert^{-\xi}, & \kappa_T &\propto \vert t \vert^{-\gamma}, & \eta &\propto \vert t \vert^{\beta}, & \vert P - P_c \vert &\propto \vert \epsilon \vert^{\delta}.
\end{align*}
When $\vert t \vert \ll 1$ and $\vert \xi \vert \ll 1$, and by making use of the outcomes presented in Section \ref{Sec:3}, we obtained
\begin{equation}\label{Eq.18}
p = 1 + u_{10} t + u_{01} \epsilon + u_{11} t \epsilon + u_{02} \epsilon^2 + u_{03} \epsilon^3 + O(t \epsilon^2, \epsilon^4),
\end{equation}
where 
\begin{align}
u_{01} &= u_{02} = 0, \\
u_{10} &= - u_{11} = \frac{T_c}{P_c v_c}, \\
u_{03} &= -\frac{40 Q^2 - 40 c^n \alpha + 40 c^n n \alpha - 2 v_c^2 + \pi T_c v_c^3}{\pi P_c v_c^4}.
\end{align}
Upon substitution of the analytical expression of the critical point into Eq. (\ref{Eq.18}), we achieve
\begin{equation}\label{Eq.22}
p = 1 + \frac{8t}{3} - \frac{8t\epsilon}{3} - \frac{4}{3} \epsilon^3 + O(t \epsilon^2, \epsilon^4),
\end{equation}
in the given equations, $\epsilon_{l}$ denotes the larger black hole, while $\epsilon_{s}$ signifies the smaller one,
\begin{align}\label{Eq.23}
2t(\epsilon_s - \epsilon_l) + (\epsilon_s^3 - \epsilon_l^3) &= 0,
\end{align}
\begin{align}\label{Eq.24}
\int_{\epsilon_s}^{\epsilon_l} \frac{\epsilon dp}{d\epsilon} d\epsilon &= 0,
\end{align}
finding solutions to these equations yields,
\begin{equation}\label{Eq.25}
\epsilon_l = -\epsilon_s = \sqrt{-2t}.
\end{equation}
From Eq. (\ref{Eq.25}), it can be observed that the critical exponent $\beta$ is equal to $\frac{1}{2}$. This leads us to the conclusion that the exponent $\gamma$ is equal to 1, based on the following equation,
\begin{equation}\label{Eq.26}
\kappa_T \propto \frac{\partial \epsilon}{\partial p} = \frac{1}{u_{11} t}.
\end{equation}
To calculate the critical exponent $\delta$, we set  $t = 0$ in Eq. (\ref{Eq.18}). This yields $p - 1 = -\frac{4}{3} \epsilon^{3}$,  resulting in $\delta = 3$.  Consequently, we conclude that the three critical exponents obtained in this study are identical to those of the RN-AdS black hole, as discussed in  Ref.~\cite{34}.

 		\section{  The modification caused by the presence of perfect fluid matter on the phase transition of AdS black holes in the context of Rastall gravity.}
 		\label{Sec:5}
This section   investigates the impact of perfect fluid matter on the phase transitions of AdS black holes within the framework of Rastall gravity,  analyzing  various critical behaviors. The following equation is derived by substituting Eqs. (\ref{Eq.6}), (\ref{Eq.7}), (\ref{Eq.8}), and (\ref{Eq.15}) into Eq. (\ref{Eq.5}) under the assumption of a small limit for $j$. The specific formulas for $T_{1}$, $T_{2}$, $T_{3}$, $T_{4}$, and $T_{5}$ can be found in the appendix.

\begin{align}\label{Eq.27}
T=\dfrac{T_1}{4 \sqrt{\pi } S}+\dfrac{24 j^2 P \pi ^{3/2} T_1+9 j^2 \pi \left(\dfrac{T_2+2 (1-3 \kappa \lambda (1+\omega ))^2 T_3}{12 S^{3/2} (1-3 \kappa \lambda (1+\omega ))^2}+T_4\right)}{T_{5}}+O(j)^{3}.
\end{align}

\begin{figure}[H]
	\centering
	\includegraphics[width=0.4\linewidth, height=5.5cm]{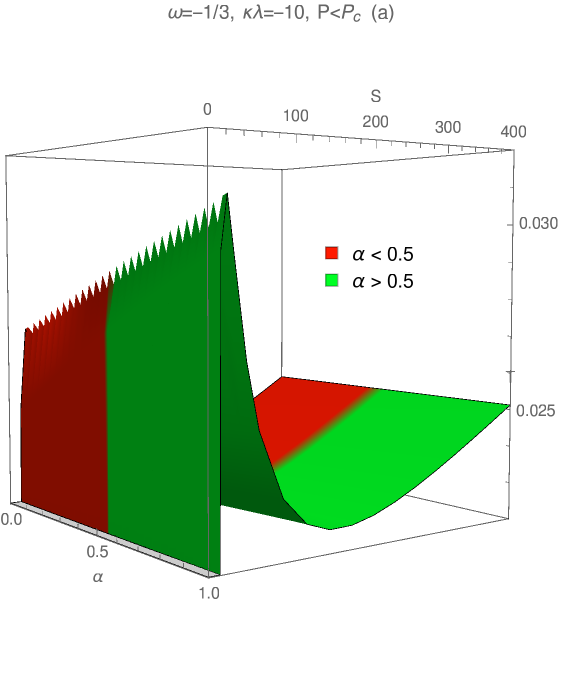}\hspace{1cm} \includegraphics[width=0.4\linewidth, height=5.5cm]{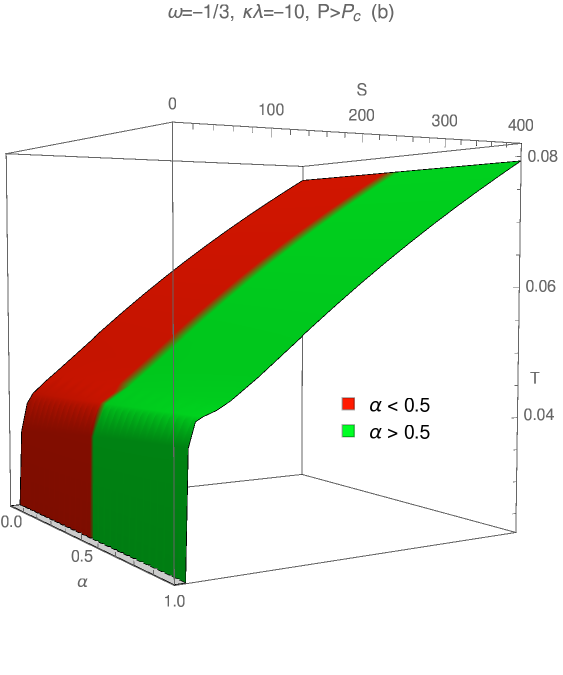}
	
	\caption{Effect of perfect fluid intensity on temperature-entropy behavior: Graphs for $P < P_{c}$ (graph (a)) and $P > P_{c}$ (graph (b), $j=Q=1$).}
	\label{fig1}
\end{figure}

In the context of Rastall gravity,  AdS black holes surrounded by perfect fluid matter  exhibits oscillatory  temperature trends when the parameter $P$ is below the critical threshold $P_{c}$. Conversely, when $P$ exceeds $P_{c}$,  the temperature graph displays a smooth profile without any extreme points, indicating the absence of oscillations. It becomes apparent that an increase in the $\alpha$ value leads to higher local maximum and minimum points on the temperature graph. As the intensity of the perfect fluid matter increases, the transition phase follows an upward trajectory.

We derive the Gibbs free energy  as a function of variables $\omega$, $\alpha$, $j$, $Q$, $S$, $\kappa\lambda$, and $P$  by substituting Eqs. (\ref{Eq.6}), (\ref{Eq.7}), (\ref{Eq.8}), and (\ref{Eq.15}) into Eq. (\ref{Eq.10}). The specific formulas for $G_{1}$, $G_{2}$, $G_{3}$, and $G_{4}$  can be found in the appendix.

\begin{align}\label{Eq.28}
G=G_1-\frac{162 j^2 \pi S \left(G_29 \sqrt{S}+\sqrt{\pi } \left(\frac{16}{9} P \left(3 \pi Q^2+S (3+8 P S)\right)+G_3\right)\right)}{9 \sqrt{S}G_4^2}-TS+O(j)^{3}.
\end{align}
 			\begin{figure}[H]
 				\centering
 				\includegraphics[width=0.4\linewidth, height=5.5cm]{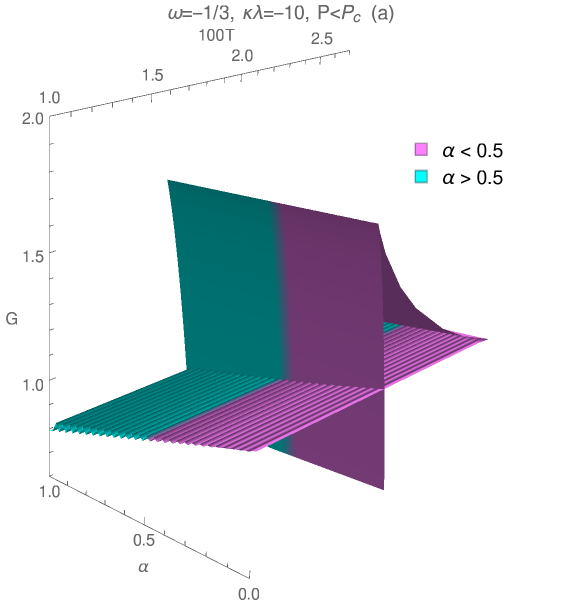}\hspace{1cm}	\includegraphics[width=0.4\linewidth, height=5.35cm]{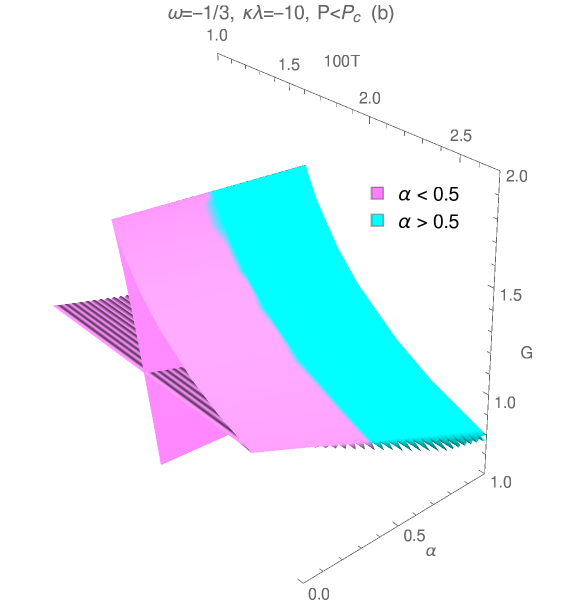} 
 				
 				\caption{ Gibbs free energy variation with temperature for various $ \alpha $ values, with $ P < P_{c} $.  }
 				\label{fig2}
 					\end{figure}
 		When $P$ is below $P_{c}$, the Gibbs free energy exhibits a pronounced swallowtail pattern. Notably, decreasing the parameter $\alpha$ shifts the intersection point in the $G$ plot towards higher $G$ values and lower $T$ values. However, in cases where $P$ surpasses $P_{c}$, the distinctive swallowtail pattern in the Gibbs free energy graph disappears entirely. It is evident that an increase in $\alpha$  leads to a rising trend for the minimal temperature.	
 				
 			To explore how the intensity of a perfect fluid  matter affects the stability of AdS black holes  in Rastall gravity, we plot the heat capacity against entropy. This is accomplished using Eq. (\ref{Eq.11}) and Eq. (\ref{Eq.27}).
 			\begin{figure}[H]
 				\centering
 				\includegraphics[width=0.4\linewidth, height=5.5cm]{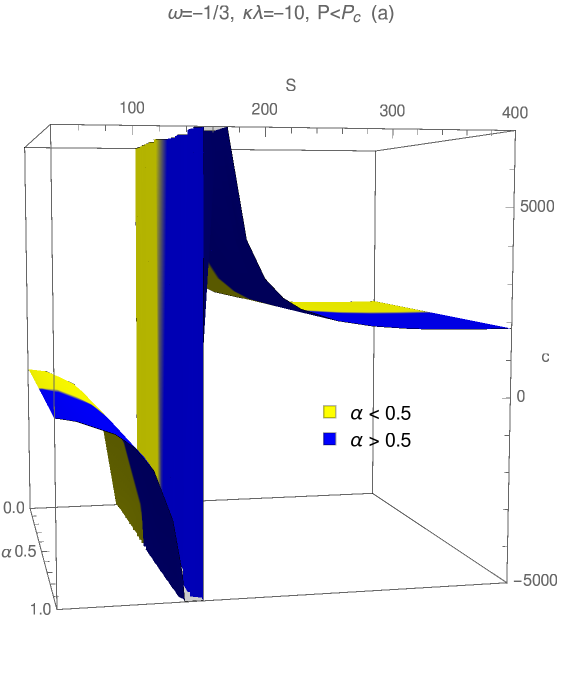}\hspace{1cm}	\includegraphics[width=0.4\linewidth, height=5.5cm]{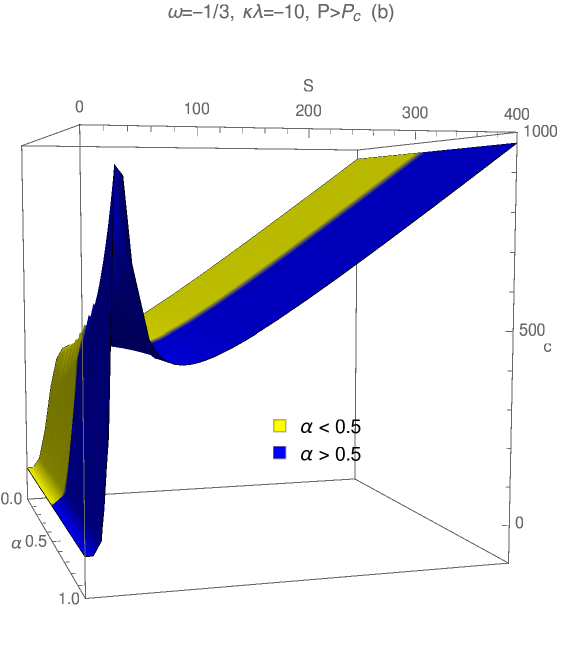} 
 				
 				\caption{ Impact of perfect fluid intensity on heat capacity trends: Plots for $ P < P_{c} $ (graph (a)) and $ P > P_{c} $ (graph (b), with  $ j $=$ Q $=1). }
 				\label{fig3}
 			\end{figure}
  The graphical representation illustrates	The response of $ C $ to different values of $ \alpha $, $ P $  and $ S $.  It reveals that  the heat capacity displays a non-continuous pattern at two specific points when $ P < P_{c} $. This trait results in a thermodynamically unstable state within the range delimited by these two divergence points. Stability is restored outside of this specific region.
 	
\begin{table}[htb]
	\centering
	\begin{tabular}{|c|c|c|c|c|c|c|c|c|c|}
		\hline
		\multicolumn{2}{|c|}{} & \multicolumn{4}{c|}{$\omega = -\frac{1}{3}$} & \multicolumn{4}{c|}{$\omega = -\frac{2}{3}$} \\ \hline
		$\kappa\lambda$ & $\alpha$ & $P_c$ & $T_c$ & $S_c$ & $G_c$ & $P_c$ & $T_c$ & $S_c$ & $G_c$ \\ \hline
		0.01 & 0.01 & 0.00180 & 0.03257 & 40.81104 & 0.96651 & 0.00179 & 0.03251 & 40.93280 & 0.92084 \\ \hline
		0.01 & 0.1 & 0.00173 & 0.03195 & 41.97380 & 0.80989 & 0.00167 & 0.03132 & 43.22592 & 0.33206 \\ \hline
		10 & 0.01 & 0.00181 & 0.03272 & 40.54781 & 0.98088 & 0.00182 & 0.03274 & 40.51990 & 0.98086 \\ \hline
		10 & 0.1 & 0.00189 & 0.03342 & 39.34339 & 0.95583 & 0.00190 & 0.03358 & 39.07093 & 0.95567 \\ \hline
		-10 & 0.01 & 0.00181 & 0.03271 & 40.57235 & 0.98081 & 0.00181 & 0.032699 & 40.59408 & 0.98066 \\ \hline
		-10 & 0.1 & 0.00187 & 0.03328 & 39.58406 & 0.95515 & 0.00186 & 0.03315 & 39.79800 & 0.95372 \\ \hline
	\end{tabular}
	\caption{Critical points for different values of $\omega$, $\alpha$, $\kappa\lambda$, and with $j = 1$, $Q = 1$.}
	\label{tt1}
\end{table}
Based on the data presented in the Table~\ref{tt1}, we observe that critical thermodynamic properties tend to   increase with an increase in the equation of state parameter $\omega$, except for the critical entropy.   In the presence of  perfect fluid   matter, there is a general decrease in critical thermodynamic values as the intensity of the perfect fluid  matter, represented by $\alpha$, increases. However, this decrease does not apply to the critical entropy. Similarly, an  increase in the Rastall gravity parameter $\left|\kappa\lambda\right|$ typically leads to higher critical values of thermodynamic properties,  with the exception of the critical entropy, which experiences a decrease.		
 				\section{Identifying fractional-order phase transitions in AdS black holes within the framework of perfect fluid matter and Rastall gravity}
 				\label{Sec:6}
 In this section, we  conducted an in-depth examination to identify fractional-order phase transitions in AdS black holes surrounded by perfect fluid matter within the framework of Rastall gravity. The equation representing the state of an AdS black hole in the presence of perfect fluid matter within the framework of Rastall gravity can be expressed as follows:		
\begin{align}\label{Eq.29}
P=\dfrac{Q^2-r_{h}^2+4 \pi  r_{h}^3 T-c^n \alpha +c^n n \alpha }{8 \pi  r_{h}^4},
\end{align}
for further examination, it is more suitable to introduce the subsequent variables with \noindent\( t  =   $ \textit{v} $  = p  = 0, \)
\begin{align}\label{Eq.30}
{P=(1+p) P_{c},\hspace{0.5cm}T=(1+t) T_{c},\hspace{0.5cm}V=(1+v) V_{c}.}
\end{align}
Through the resolution of this equation, we deduce the formulation for the variable \textit{$ v $},
 \begin{align}\label{Eq.31}
\dfrac{-8 t (1+v)^3+3 p (1+v)^4+v^3 (4+3 v)}{(1+v) \left(Q^2+c^n (-1+n) \alpha \right)}=0.
 \end{align}
 Utilizing the critical point expression derived in Section \ref{Sec:3} for charged AdS black holes in the presence of perfect fluid matter within Rastall gravity, we obtain the following expression for the rescaled Gibbs free energy,
\begin{align}\label{Eq.32}
g(t,p)&=\left(12 c^{1+n} (-1+n) \alpha -12 \sqrt{6} c^n n (1+v) \alpha  \sqrt{Q^2+c^n (-1+n) \alpha }+12 c \left(Q^2+(1+v)^2 \left(3-2
	v\right.\right.\right.\notag\\
&\left.\left.\left.+v^2-4 t (1+v)+p (1+v)^2\right) \left(Q^2+c^n (-1+n) \alpha \right)\right)\right)/\left(24 \sqrt{6} c (1+v) \sqrt{Q^2+c^n (-1+n)
		\alpha }\right).
\end{align}
The equation mentioned above can be  expressed more appropriately through the Taylor series expansion of $t$,  leading to a simpler and more concise representation.
\begin{align}\label{Eq.33}
g(t,p)= \gamma_{1}(p) + \gamma_{2}(p)t + \gamma_{3}(p)t^2 +O[t]^3,
\end{align}
where $ \gamma_{1}(p) $, $ \gamma_{2}(p) $,  and $ \gamma_{3}(p) $ represent a Puiseux series of $ p $ with fractional exponents. A significant thermodynamic quantity  examined in this work is the heat capacity, which  serves as an indicator of local thermodynamic stability. Near the critical point, the heat capacity at constant pressure, $ C_{p} $ is directly   linked to  $ \dfrac{\partial ^{2}g}{\partial t^{2}}|p $. Thus, taking into account Eq. (\ref{Eq.33}), the coefficient $ \gamma_{3}(p) $ should be proportional to $ C_{p} $. Typically, the coefficient $ \gamma_{3}(p) $ diverges at the critical point, leading to a divergent $ C_{p} $. This divergence in $ C_{p} $ indicates a second-order phase transition according to the conventional  Ehrenfest classification.\

We utilize the derived expansion series of $ \textit{v}(t, p) $ from Eq. (\ref{Eq.31}) and incorporate it into the expression for $ g(t, p) $. The resulting form is as follows
\begin{align}\label{Eq.34}
g(t,p)&=\frac{12 c^{1+n} (-1+n) \alpha -12 \sqrt{6} c^n n \alpha  \sqrt{Q^2+c^n (-1+n) \alpha }+12 c \left(Q^2+(3+p) \left(Q^2-c^n \alpha
		+c^n n \alpha \right)\right)}{24 \sqrt{6} c \sqrt{Q^2+c^n (-1+n) \alpha }}\notag\\
&-\sqrt{\frac{2}{3}} t \sqrt{Q^2+c^n (-1+n) \alpha }-\frac{4\text{  }2^{5/6}\text{  }t^2 \sqrt{Q^2-c^n \alpha +c^n n \alpha }}{9\ 3^{1/6} p^{2/3}}+O(t^{3}).
\end{align}
The representation of the equation of state in the vicinity of the critical point is as follows\

$ p=k t $, \hspace{0.5cm}{with} $ k=\dfrac{T_c}{P_c V_c}=\dfrac{8}{3}$.\

In this context, we adopt the Caputo definition, which facilitates the convenient application of standard boundary and initial conditions~\cite{72}.
\begin{align}\label{Eq.35}
D_t^{\beta }g(t)=\dfrac{1}{\Gamma (n-\beta )}\int _0^t(t-\tau )^{n-\beta -1}\dfrac{\partial ^ng(\tau )}{\partial \tau ^n}\text{d$\tau
		$},\hspace{2cm} n-1<\beta <n,
	\end{align}
	When conducting the computation, considering $ \beta $ as the order of derivative and taking the limit as both $ t $ and $ p $ approach zero, the result is as follows:
	\begin{align}\label{Eq.36}
D_t^{\beta }g(t,p)|_p= \dfrac{2 t^{2-\beta }}{\Gamma (3-\beta )}\gamma_{3}(p),\text{   }t>0;1<\beta <2,
	\end{align}
	$$ \lim_{t \to 0} D_t^{\beta }g(t,p)|_p  =\begin{cases}
	0 \hspace{1cm}for \hspace{0.2cm} \beta < \frac {4}{3},  \\ {\text{}_+^- } {{\dfrac{2^{5/6} t^{\frac{4}{3}-\beta } \sqrt{Q^2+c^n (-1+n) \alpha }}{\sqrt{3} \Gamma \left[\frac{2}{3}\right]}}}  \hspace{1cm}for \hspace{0.2cm}  \beta = \frac {4}{3}, \\ {\text{}_+^-\infty }\hspace{1cm}for\hspace{0.2cm}\beta > \frac {4}{3}.
	\end{cases}$$
	
The study results offer compelling proof that within the particular context being examined, encompassing a spherically symmetric Anti-de Sitter (AdS) black hole amid a perfect fluid matter  environment and a Rastall gravity, the characteristics of the fractional phase transition persist unaltered and consistently align with a 4/3 order.

\section{Conclusions}
\label{Sec:7}
In this paper, we  examined the phase transition of AdS black holes  within the Rastall gravitational   framework  with existence  of a perfect fluid.  Compared to previous work,  remarkable progress in this study was highlighted  by  determining analytical expressions for critical points for charged and Kerr-Newman AdS black holes using an approximated expression for the horizon radius. A new viable condition was discovered  through the analysis of the new expressions of critical points, which  established a relationship between the Rastall parameter $\kappa\lambda$ and the equation of state parameter $\omega$ as  follows: $\kappa\lambda = \frac{\omega}{1 + \omega}$.
 Due to this new condition,  we were  able to reproduce all the analytical expressions of critical points calculated within the framework of Einstein's general relativity,  in the  two cases: a charged AdS black hole and a rotating AdS black hole. These results suggest that Rastall gravity could serve as an alternative theory of gravitation to general relativity.

An examination of the  phase diagrams showed that the position of the critical point is  highly influenced by the  presence of perfect fluid matter around Anti-de Sitter (AdS) black holes, although it does not  change the  manifestation of phase transitions between small and large black holes. The graphical analysis of the temperature $T$ reveals intriguing patterns. When $\alpha$ is changed while keeping $\omega$ and $\kappa\lambda$ constant, the local maxima and minima of $T$ increase with increasing values of $\alpha$. Moving on to the Gibbs free energy $G$ graph, the impact of the intensity of perfect fluid matter $\alpha$ on the swallowtail pattern of $G$ becomes evident. As $\alpha$ increases while keeping $\omega$ and $\kappa\lambda$ constant, the intersection point of $G$ shifts towards higher $T$ values and lower $G$ values. The graphical representation of the heat capacity's response to various values of $\alpha$, $P$, and $S$ reveals a discontinuity at two specific points. Stability is restored beyond this particular region.

The survey results  provide  strong evidence suggesting that, in the    given  context  with investigation featuring a spherically symmetric Anti-de Sitter (AdS) black hole within a perfect fluid matter environment and subject to Rastall gravity, the characteristics of the fractional phase transition   remain unaltered. Additionally, these features consistently adhere to a 4/3 order, highlighting the stability and consistency of the observed phenomena. Furthermore, the investigation extends to computing critical exponents, offering valuable insights into the behavior of relevant thermodynamic quantities near inflection points.
\section*{Acknowledgements}
The authors would like to thank the anonymous referee for interesting comments and suggestions which motivated us to prepare a well-improved revised version.

	\newpage
\appendix
\textbf{\large{Appendix}}\

The Hawking temperature for Kerr-Newman Anti-de Sitter (Ads) black holes, considering Rastall gravity and the presence of a perfect fluid matter, is denoted by the expressions of terms $  T_{1}, T_{2}, T_{3}, T_{4} $, and $ T_{5} $ , which are presented as follows
 		 \begin{align}
T_{1}=& 3 \pi ^{\dfrac{-1-3 \omega +6 \kappa \lambda  (1+\omega )}{-2+6 \kappa \lambda  (1+\omega )}} S^{\dfrac{-1+3 \omega }{-2+6 \kappa
		\lambda  (1+\omega )}} \alpha  (\kappa \lambda -\omega +\kappa \lambda  \omega )+\left(-\pi  Q^2+S+8 P S^2\right) \notag\\
&(-1+3 \kappa \lambda  (1+\omega ))\left/\left( S^{1/2} (-1+3 \kappa \lambda  (1+\omega ))\right)\right.,\notag
\end{align}
\begin{align}  T_2 &=-6 \pi ^{\dfrac{-2-3 \omega +9 \kappa \lambda  (1+\omega )}{-2+6 \kappa \lambda  (1+\omega )}} S^{\dfrac{-1+3 \omega }{-2+6 \kappa
		\lambda  (1+\omega )}} (3+8 P S) \alpha  (-1+3 \omega ) (\kappa \lambda -\omega +\kappa \lambda  \omega ),\notag\end{align}		
\begin{align}
T_{3}=&-\pi ^{3/2} Q^2 (3+8 P S)+\pi ^{\dfrac{-2-3 \omega +9 \kappa \lambda  (1+\omega )}{-2+6 \kappa \lambda  (1+\omega )}} S^{\dfrac{-1+3
		\omega }{-2+6 \kappa \lambda  (1+\omega )}} (3+8 P S) \alpha -\dfrac{1}{3} \sqrt{\pi }\notag\\
& S \left(27+288 P S+832 P^2 S^2\right),\notag\end{align}
\begin{align}
T_4 =&\left(9 \pi ^{1+\dfrac{1-3 \omega }{-2+6 \kappa \lambda  (1+\omega )}} S^{-1+\dfrac{-1+3 \omega }{-2+6 \kappa \lambda
		(1+\omega )}} \left(-1-\frac{8 P S}{3}\right) \alpha  (-1+3 \omega )+16 P \left(3 \pi  \right.\right.\notag\\
&Q^2+S (3+8 P S)) (-2+6 \kappa \lambda  (1+\omega )))\sqrt{\pi }\left/\left(9 \sqrt{S} (-2+6 \kappa \lambda  (1+\omega ))\right)\right.,\notag\end{align}
\begin{align}
T_{5}={\left(3 \pi  Q^2+3 S+8 P S^2-3 \pi ^{1+\dfrac{1-3 \omega }{-2+6 \kappa \lambda  (1+\omega
			)}} S^{\dfrac{-1+3 \omega }{-2+6 \kappa \lambda  (1+\omega )}} \alpha \right)^2}.\notag
\end{align}
The Gibbs free energy components $ G_{1},  G_{2},G_{3} $, and $ G_{4} $ are expressed as follows 
	\begin{align}
{G_1=\dfrac{8 P S^2+3 \left(\pi  Q^2+S\right)-3 \pi ^{1+\dfrac{1-3 \omega }{-2+6 \kappa \lambda  (1+\omega )}} S^{\dfrac{-1+3 \omega
				}{-2+6 \kappa \lambda  (1+\omega )}} \alpha }{6 \sqrt{\pi } \sqrt{S}}},\notag
\end{align}

\begin{align}
{G_2=\dfrac{\sqrt{\pi } (3+40 P S) \left(-3 \pi  Q^2-S (3+8 P S)+3 \pi ^{1+\dfrac{1-3 \omega }{-2+6 \kappa \lambda  (1+\omega )}} S^{\dfrac{-1+3
				\omega }{-2+6 \kappa \lambda  (1+\omega )}} \alpha \right)}{162 S^{3/2}}},\notag\end{align}
\begin{align}
{G_3=\dfrac{\pi ^{1+\dfrac{1-3 \omega }{-2+6 \kappa \lambda  (1+\omega )}} S^{-1+\dfrac{-1+3 \omega }{-2+6 \kappa \lambda  (1+\omega
				)}} (-3-8 P S) \alpha  (-1+3 \omega )}{3 (-2+6 \kappa \lambda  (1+\omega ))}},\notag
\end{align}

\begin{align}
{G_4=\left(-3 \pi  Q^2-S (3+8 P S)+3 \pi ^{1+\dfrac{1-3 \omega }{-2+6 \kappa \lambda  (1+\omega )}} S^{\dfrac{-1+3 \omega }{-2+6 \kappa
			\lambda  (1+\omega )}} \alpha \right)}.\notag
\end{align}
\end{document}